\newfam\amsfam
\font\tenams=msam10 \textfont\amsfam=\tenams
\mathchardef\ls="382E\mathchardef\gs="3826
\input epsf
\def\a{\alpha}\def\d{\delta}\def\e{\epsilon}
\def\f{\phi}\def\g{\gamma}\def\h{\theta}
\def\la{\lambda}\def\m{\mu}\def\n{\nu}\def
\p{\pi}\def\r{\rho}\def\s{\sigma}

\def\Om{\Omega}\def\D{\Delta}

\def\de{\partial}
\def\inf{\infty}\def\id{\equiv}\def\mo{{-1}}\def\ha{{1\over 2}}

\def\({\left(}\def\){\right)}\def\[{\left[}\def\]{\right]}
\def\mn{{\mu\nu}}\def\ij{{ij}}

\def\af{asymptotically flat }
\def\fe{field equations }

\def\cc{coupling constant }
\def\ssy{spherically symmetric }\def\bc{boundary conditions }
\def\RN{Reissner-Nordstr\"om }\def\GB{Gauss-Bonnet }
\def\ab{asymptotic behavior }
\def\wrt{with respect to }

\def\section#1{\bigskip\noindent{\bf#1}\smallskip}

\font\smalll = cmr7

\def\PL#1{Phys.\ Lett.\ {\bf#1}}\def\CMP#1{Commun.\ Math.\ Phys.\ {\bf#1}}

\def\PR#1{Phys.\ Rev.\ {\bf#1}}\def\CQG#1{Class.\ Quantum Grav.\ {\bf#1}}
\def\NP#1{Nucl.\ Phys.\ {\bf#1}}
\def\JMP#1{J.\ Math.\ Phys.\ {\bf#1}}
\def\PRS#1{Proc.\ R. Soc.\ Lond.\ {\bf#1}}
\def\JoP#1{J.\ Phys.\ {\bf#1}} \def\IJMP#1{Int.\ J. Mod.\ Phys.\ {\bf #1}}

\def\AoP#1{Ann.\ Phys.\ {\bf#1}}

\def\arx#1{{\tt arXiv:#1}}

\def\ref#1{\medskip\everypar={\hangindent 2\parindent}#1}
\def\beginref{\begingroup
\bigskip
\centerline{\bf References}
\nobreak\noindent}

\def\at{{\tilde\a}}\def\rp{{r^\ast_+}}\def\rpm{{r^\ast_\pm}}
\def\en{e^\n}\def\el{e^\la}

{\nopagenumbers
\line{}
\vskip40pt
\centerline{\bf Dyonic black holes in Kaluza-Klein theory with a Gauss-Bonnet action}
\vskip50pt

\centerline{{\bf S. Mignemi}\footnote{$^\ast$}{e-mail: smignemi@unica.it}}
\vskip5pt
\centerline {Dipartimento di Matematica, Universit\`a di Cagliari}
\centerline{via Ospedale 72, 09124 Cagliari, Italy}
\smallskip
\centerline{and INFN, Sezione di Cagliari}
\centerline{Cittadella Universitaria, 09042 Monserrato, Italy}

\vskip40pt
\centerline{\bf Abstract}
\medskip
{\noindent We consider a five-dimensional Einstein-Gauss-Bonnet model, which gives rise after dimensional reduction
to Einstein gravity nonminimally coupled to nonlinear electrodynamics. The black hole solutions of the four-dimensional
model modify the \RN solutions of general relativity. The gravitational field presents the standard singularity at $r=0$,
while the electric field can be regular everywhere if the magnetic charge vanishes.}
\vskip60pt
\vfil\eject}

\section{1. Introduction}
In his career, Richard Kerner has given important contributions to the Kaluza-Klein framework, in particular showing
that also nonabelian gauge theories can be obtained from the process of dimensional reduction by increasing
the number of internal dimensions [1].
He was also one of the first to notice the relevance of \GB (GB) terms in the action of higher-dimensional theories,
showing that they give rise to nonlinear contributions to the electrodynamics in the reduced theory [2].\footnote{$^1$}
{This was also independently observed by Buchdal [3].}

Nonlinear electrodynamics was first proposed by  Heisenberg and Euler [4] to give an effective classical description of
quantum electrodynamics in a suitable limit.
A general formulation, which includes also the model studied in [2.3] as a special case, was given by Plebanski [5].
However, the action obtained from dimensional reduction of the GB action enjoys peculiar algebraic properties.
Some solutions of this model in flat space, hence neglecting gravity, have been discussed in [2,6,7].

In the present paper, we are interested in the full theory containing both gravity and Maxwell fields.
We have recently shown that a five-dimensional Kaluza-Klein theory containing GB contributions
admits exact solutions that modify the \RN (RN) metric of general relativity [7]. In particular, when the nonminimal coupling between
gravity and Maxwell fields arising from the dimensional reduction is neglected, its dyonic solutions display an everywhere regular
electric field [7]. These modifications could in principle give experimental evidence of the existence of extra dimensions.

Here, we give a more complete discussion of the dyonic solutions of the model,
that takes into account also the nonminimal interaction terms. It turns out that, contrary to the case where these terms
are neglected, the electric field can be regular everywhere only for vanishing magnetic field.\footnote{$^2$}{The existence of
solutions with regular electric field was predicted also in [8], using different methods.} This may be considered as a favourable
feature of the model, since magnetic monopoles are not observed in nature.

\section{2. The model}
As is well known, the Einstein-Hilbert action can be generalized in dimensions higher than four, by the introduction
of the Lovelock terms [9]. These give the most general extensions of the Einstein-Hilbert action
that give rise to second order field equations in arbitrary dimensions.
One of their most notable properties is that they do not introduce new degrees of freedom
in the spectrum in addition to the graviton, and therefore avoid the presence of ghosts or tachyons, in contrast
with most higher-derivative actions [10]. In lower dimensions they are total derivatives and do not contribute to the equations
of motion.
In particular, in five dimension, the only term of this type is the so-called Gauss-Bonnet term,
$$S=R^{\m\n\r\s}R_{\m\n\r\s}-4R^{\m\n}R_{\m\n}+R^2.$$

The dimensional reduction of these generalized actions gives rise to models of gravity coupled to nonlinear electrodynamics,
which as a consequence of the properties of the higher-dimensional theory, contain only graviton and photon excitations,
and are therefore relevant from a phenomenological perspective.

Therefore, we consider a five-dimensional Einstein-Gauss-Bonnet theory, with action [1,2,7]
$$I=\int\sqrt{-g}\ d^5x(R+\a S),\eqno(1)$$
where $\a$ is a coupling constant  and $R$ the Ricci scalar.
The \cc $\a$ has dimension $[L]^2$, and is usually  assumed to be positive for stability reasons.
Arguments based on quantum gravity or string theory fix it to be of Planck scale, but in any case observations set a very small
upper limit on its value [8].

We make the simple ansatz for the five-dimensional metric\footnote{$^3$}{We set $\m,\n=0,\dots,4$; $i,j=0,\dots,3$.} [9-10]
$$g_\mn=\(\matrix{g_\ij+4A_iA_j&2A_i\cr2A_j&1}\),\eqno(2)$$
where $A_i$ is the Maxwell potential and we have chosen the normalization in order to simplify the dimensionally
reduced action.

Discarding total derivatives, the action reduces to [2,3,13]
$$I=\int\sqrt{-g}\ d^4x\[R-F^\ij F_\ij+3\a\Big[(F^\ij F_\ij)^2-2F^{ij}F_{jk}F^{kl}F_{li}\Big]-2\a L_{int}\],\eqno(3)$$
where $F_\ij=\de_iA_j-\de_jA_i$ and the interaction term is
$$L_{int}=F^\ij F^{kl}(R_{ijkl}-4R_{ik}\d_{jl}+R\,\d_{ik}\d_{jl}).\eqno(4)$$

If one neglects $L_{int}$, the action describes a model of gravity minimally coupled to a specific form of nonlinear electrodynamics.
Exact \af \ssy solutions of (3) in the absence of $L_{int}$ have been investigated in [7], where it was shown that the \RN solution of
general relativity is modified in the dyonic case.
In particular, for $\a>0$, the solutions read
$$A=a(r)\,dt+ P\cos\h\,d\f,\eqno(5)$$
$$ds^2=-e^{2\n}dt^2+e^{-2\n}dr^2+r^2(d\h^2+\sin^2\h\, d\f^2),\eqno(6)$$
with $P$ the magnetic monopole charge, and radial electric field $E\id F_{01}$,
$$E={da\over dr}={Qr^2\over r^4+3\a P^2},\eqno(7)$$
while the metric function is
$$e^{2\n}=\ 1-{2M\over r}+{P^2\over r^2}+{Q^2\over2\sqrt{2\at}r}\[\p+\arctan\(1-{\sqrt2\,r\over\sqrt\at}\)-\arctan\(1+{\sqrt2\,r\over\sqrt\at}\)
+\ha\log{r^2-\sqrt{2\at}\,r+\at\over r^2+\sqrt{2\at}\,r+\at}\],\eqno(8)$$
where we have set $\at=\sqrt{3\a P^2}$.
The \ab of the solution is given by
$$e^{2\n}=1-{2M\over r}+{Q^2+P^2\over r^2}-{3\bar\a P^2Q^2\over5r^6}+o\({1\over r^7}\),\eqno(9)$$
and has therefore the same form as for RN up to order $1/r^5$.
It follows that the integration constants $M$ and $Q$ can be identified with the mass and the electric charge.
The electric field is regular everywhere, and the properties of the metric are analogous to those of
the RN solution: a curvature singularity is present at the origin and for $M$ greater than its extremal value is shielded by two horizons.

If $\a<0$, instead,  a singularity of the electric field appears at $r_0=\sqrt\at$, where now  $\at=\sqrt{-3\a P^2}$.
The metric takes a slightly different form,
$$e^{2\n}=\ 1-{2M\over r}+{P^2\over r^2}+{Q^2\over2\sqrt\at\,r}\[{\p\over2}-\arctan{r\over\sqrt\at}-\ha\log{r-\sqrt\at\over r+\sqrt\at}\].\eqno(10)$$

A spherical curvature singularity occurs at $r=r_0$ and the solution can have one or two horizons depending on the specific values of the parameters,
while the \ab is still given by (9).

It is evident that the effects of nonlinear electrodynamics are more relevant at small $r$, and tend to vanish at infinity.

\section{3. The solution}
We now consider the \af\ssy solution of the \fe stemming from the action (3), when also the term $L_{int}$ is included.
In this case it is not possible to find an exact solution, and we must proceed perturbatively.

We look for spherically symmetric solutions with electromagnetic field of the dyonic form (5) and
$$ds^2=-e^{2\n}dt^2+e^{2\la-2\n}dr^2+e^{2\r}d\Om^2,\eqno(11)$$
where $\n$, $\la$ and $\r$ are functions of $r$.

After integration by parts, the action becomes
$$\eqalign{I=&\ 2\int dr\bigg[(2\n'\r'+\r'^2)e^{2\n-\la+2\r}+e^{\la}+a'^2e^{-\la+2\r}-P^2e^{\la-2\r}\cr
&+4\a\bigg(3P^2a'^2e^{-\la-2\r}+a'^2\r'^2e^{2\n-3\la+2\r}-a'^2e^{-\la}+2P^2\n'\r'e^{2\n-\la-2\r}\bigg)\bigg]},\eqno(12)$$
where $'=d/dr$.

Varying \wrt the fields, and then choosing the gauge $e^\r=r$, the independent \fe can be put in the form
$$\eqalignno{&\la'=-{4\a\over r}\(a'^2e^{-2\la}+{P^2\over r^3}\la'+{3P^2\over r^4}\),&(13)\cr
&(re^{2\n})'+4\a P^2{(e^{2\n})'\over r^3}=\(1-{P^2\over r^2}\)e^{2\la}-r^2a'^2+4\a\,a'^2\(1-3e^{2\n-2\la}-{3P^2\over r^2}\),&(14)\cr
&\[r^2e^{-\la}\(1+12\a\,{P^2\over r^4}+4\a\,{e^{2\n-2\la}-1\over r^2}\)a'\]'=0.&(15)}$$
An important effect of the nonminimal gravity-Maxwell coupling $L_{int}$ is that the metric field $\la$ no longer vanishes.
This is a common feature in presence of nonminimally coupled Maxwell fields.

Also the radial electric field $E\id F_{01}$ is modified \wrt the solution (7), since (15) gives
$$E=a'={Q\,r^2e^{\la}\over r^4+4\a\(e^{2\n-2\la}-1\)r^2+12\a P^2},\eqno(16)$$
where $Q$ is an integration constant that can be identified with the electric charge.

The equations (13)-(15) do not admit a solution in analytical form, so we perturb in the small parameter $\a$ around the \RN background
$$ds^2=-Adt^2+A^\mo dr^2+r^2d\Om^2,\qquad E={Q\over r^2},\eqno(17)$$
with
$$A=1-{2M\over r}+{Q^2+P^2\over r^2}.\eqno(18)$$
The perturbative expansion will be valid for large values of $r$, namely $r\gg\sqrt\a$.

As is well known, the horizons of the RN metric are located at
$$r^\ast_\pm=M\pm\sqrt{M^2-Q^2-P^2}.\eqno(19)$$
In order to simplify some expressions, in the following we shall write the RN mass $M$ in terms of the outer horizon and of the charges as
$$M={{r^\ast_+}^2+Q^2+P^2\over2r^\ast_+}.\eqno(20)$$

We now define the perturbations $\s(r)$, $\g(r)$ and $\f(r)$ through an expansion at order $\a$ around a RN background,
$$e^{2\n}=A+\a\s,\qquad\la=\a\g,\qquad E={Q\over r^2}(1+\a\f),\eqno(21)$$
 where of course, $\la=0$ for RN.

Integrating (13), we obtain at order $\a$
$$\g={Q^2+3P^2\over r^4}.\eqno(22)$$
Substitution in (15) gives
$$\f={8M\over r^3}-{3Q^2+13P^2\over r^4}.\eqno(23)$$
Finally, substituting the previous results in (13) and integrating, one obtains
$$\s={2(Q^2-P^2)\over r^4}-{2M(Q^2-P^2)\over r^5}+{6Q^4-8P^2Q^2-2P^4\over5r^6}.\eqno(24)$$

In all the solutions, we have chosen \bc such that the corrections vanish at infinity.
Hence, at order $\a$,
$$e^{2\n}\sim1-{2M\over r}+{Q^2+P^2\over r^2}+2\a\({Q^2-P^2\over r^4}-{M(Q^2-P^2)\over r^5}+{3Q^4-4P^2Q^2-P^4\over5r^6}\),\eqno(25)$$

$$e^{2\la}\sim1+2\a\,{Q^2+3P^2\over r^4},\eqno(26)$$

$$E\sim{Q\over r^2}\(1+{8\a M\over r^3}-{\a(3Q^2+13P^2)\over r^4}\).\eqno(27)$$
Our approximation works well for $r\to\inf$. At leading orders in $1/r$ the \ab is the same as in RN. Therefore, we can still identify
$M$ with the mass of the black hole and $Q$ and $P$ with its electric and magnetic charge.
Notice that the corrections to the RN solutions are much larger than in the case where the $L_{int}$ term is neglected,
since they are now $o(1/r^4)$.
Also, they are no longer symmetric in $Q$ and $P$ at leading order in $\a$.

The horizons are displaced with respect to the RN solutions, where they are located at $r^\ast_\pm$.
At first order in $\a$, one has $r_\pm=r^\ast_\pm+\a\D r_\pm$, where
$$\D r_\pm=-{\s\over A'}\Big|_{r=r^\ast_\pm}.\eqno(28)$$
It follows that
$$\D r_\pm={5(P^2-Q^2)\rpm^2-3 P^4 + 8 P^2 Q^2 - Q^4\over5\rpm^3 (\rpm^2-P^2 - Q^2)},\eqno(29)$$
where to simplify the expression we have chosen as independent parameters $r^\ast_+\,(r^\ast_-)$, $Q$ and $P$.
The actual values of these displacements strongly depend on the  charges.

A calculation shows that the condition of extremality is, at first order in $\a$,
$$M^2=P^2+Q^2+{\a\over5}\,{P^4+4Q^2P^2-3Q^4\over (P^2+Q^2)^2}.\eqno(30)$$
Depending on the values of $Q$ and $P$ the correction with respect to the RN case can be both positive or negative.
We recall however that, while the value of $r_+$ obtained in this way is in general well approximated by (29), the value of $r_-$  is reliable
only for very small values of $\a$.

In this approximation, the metric function $e^{2\n}$ does not differ much from that of RN and the causal structure should therefore be analog.
Hence, for $M$ greater than extremality, one has two horizon, while a naked singularity is present for $M$ less than its extremal value.
However, this is not necessarily true for greater values of $\a$, where the approximation fails.
\medskip

Using standard definitions it is possible to derive the thermodynamical quantities associated to the black hole
from the behaviour of the metric functions near the outer horizon,.
The temperature can be calculated from the formula
$$T={1\over4\p}\,e^{-\la}(e^{2\n})'\big|_{r=r_+}\sim {1\over4\p}\Big[A'(1-\a\g)+\a\s'\Big]_{r=r_+}.\eqno(31)$$
Hence,
$$\eqalign{ T\sim&\ {1\over4\p\rp^3}\bigg(\rp^2-P^2-Q^2\cr
&-2\a\ {7 P^6 + 25 P^4 Q^2 + 17 P^2 Q^4 - Q^6 - 14 P^4\rp^2 -
36 P^2 Q^2\rp^2 + 2 Q^4\rp^2 + 5(Q^2+P^2)\rp^4\over 5\rp^4(\rp^2-P^2-Q^2)}\bigg).}\eqno(32)$$

The entropy $S$ is usually identified with the area of the horizon, namely,
$$S\sim4\p\rp^2\(1-2\a{3P^4-8P^2Q^2+Q^4+5(Q^2-P^2)\rp^2\over5\rp^4(\rp^2-P^2-Q^2)}\).\eqno(33)$$
It follows that the thermodynamical quantities display a complicate dependence on the charges.

It is also interesting to investigate the behavior of the solutions near the singularity. This can be calculated by an expansion
in powers of $r$ near $r=0$. Setting
$$e^\n\sim r^h,\qquad e^\la\sim r^l,\qquad E\sim r^k,$$
and substituting in the \fe (13)-(15), one obtains $h=-2$, $l=-3$ and $k=-1$.
It follows that the metric functions and the electric field diverge for $r=0$, thus destroying the nice property of the solution
of sect.~2 to have a finite electric field at the origin.

The only exception is for $P=0$. In this case, the electric field vanishes at the origin.
In fact now $h=-1$, $l=0$ and $k=1$. The existence of electric solutions regular at the origin in absence of magnetic field has  also
been noticed in [8]. However, for small values of $Q$, numerical solutions show that the solutions become singular at a point $r_0>0$,
presenting a spherical singularity, similarly with the $\a<0$ solutions of sect.~2.

Remarkably, the behaviour of the solution is therefore  opposite to those with $L_{int}=0$, since regular solution can now exist only if $P=0$.

\section{4. Numerical calculations}

The solutions of (13)-(15) can also be obtained numerically. This is especially interesting for  $r\ll\a$, where the
perturbative calculation of the previous section fails. In fig.~1 are reported the metric functions and the electric field
for $\a=0.01$, $M=1$, and several values of $Q$ and $P$, such that $Q^2+P^2=1/2$, so that $r_+\sim0.7$.
The metric functions $\en$ and $\el$ do not change much for different values of the charges. In particular, $\e^{2\n}$ is similar to
the RN solution, while $e^{-2\la}=1$ for $r\gg\sqrt\a$  and then fades to 0  for $r\to0$.
In general, a curvature singularity is present at $r=0$ and the causal structure is essentially the same as that of the RN solution.
Also the electric field  is singular at the origin  as in the RN solution.

As mentioned before, an interesting special case is given by $P=0$. In fig.~2 are depicted the metric functions $e^{2\n}$, $e^{-2\la}$
and the electric field $F$ for $M=1$ and different values of the electric charge. For our choice of the parameters, if $Q>0.44$
the electric field is regular at the origin. The possibility of such behaviour had been noticed in [8] using different methods.
However, for smaller $Q$ a singularity occurs for a finite value of $r$ and the metric functions and the electric field diverge there.
\medskip
\vfil\eject
\centerline{\epsfysize=4.5truecm\epsfbox{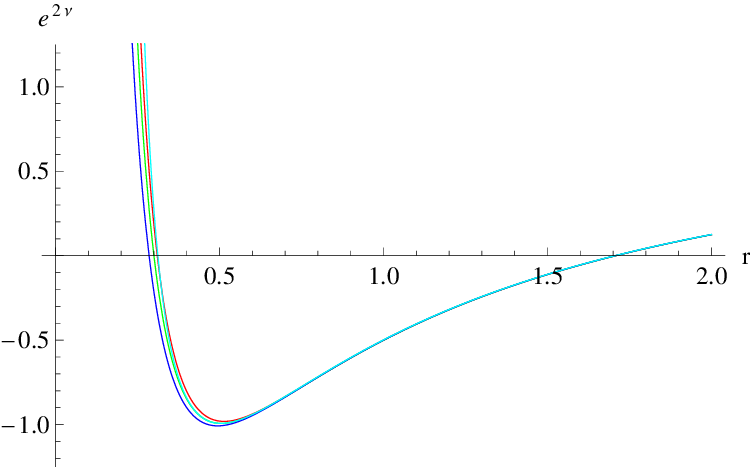}\qquad\epsfysize=4.5truecm\epsfbox{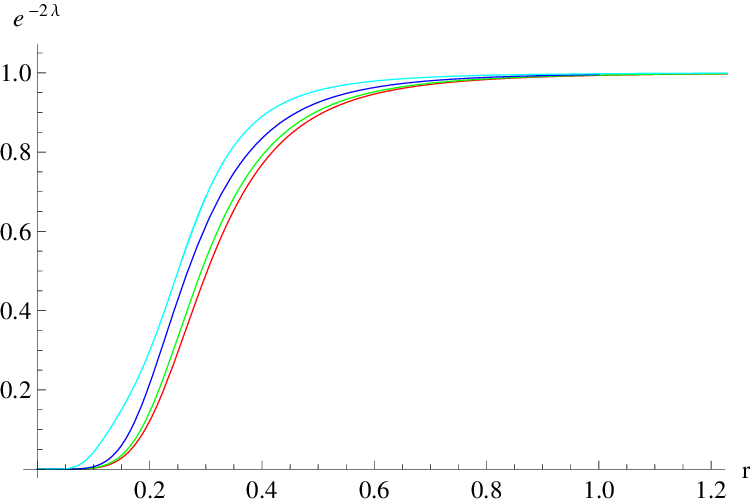}}
\medskip
\centerline{\epsfysize=4.5truecm\epsfbox{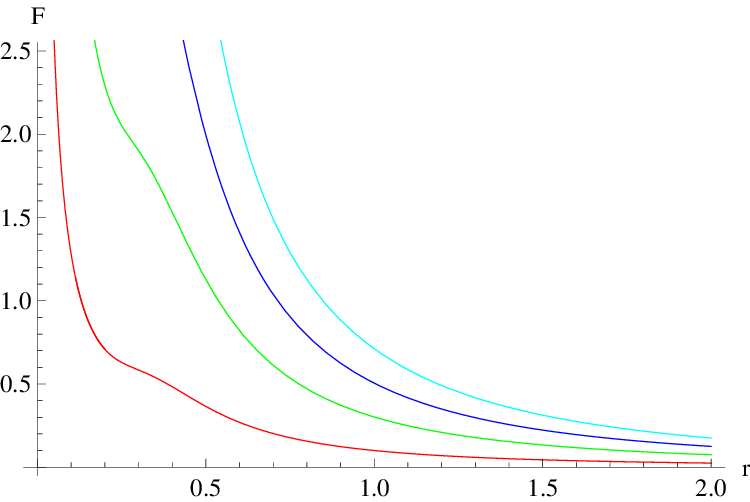}}
\medskip
\baselineskip10pt{\noindent{\smalll Fig.\ 1: The metric function $\scriptstyle{e^{2\n}}$ and  $\scriptstyle{e^{-2\la}}$ and the
electric field $\scriptstyle{F}$ for black holes
with mass $\scriptstyle{M\;=\;1}$, $\scriptstyle{Q^2+P^2\;=\;{1\over2}}$ and $\scriptstyle{Q\;=\;0.1}$ (in cyano), $\scriptstyle{Q\;=\;0.3}$ (in blue),
$\scriptstyle{Q\;=\;0.5}$ (in green), $\scriptstyle{Q\;=\;0.7}$ (in red).}
\bigskip
\baselineskip12pt
\medskip
\centerline{\epsfysize=4.5truecm\epsfbox{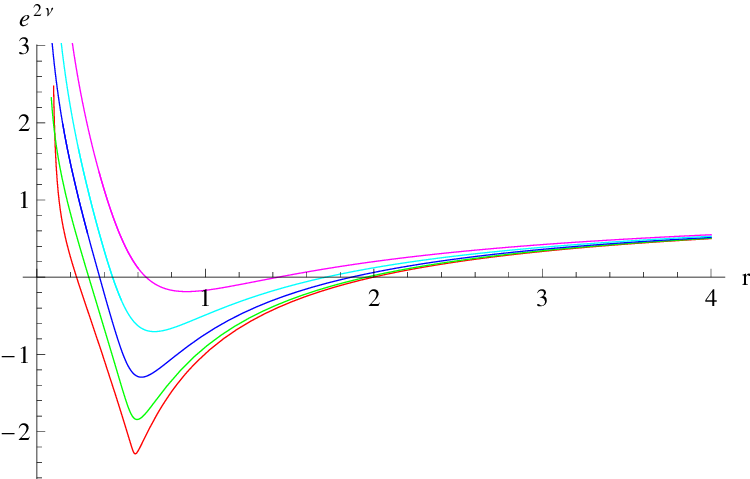}\qquad\epsfysize=4.5truecm\epsfbox{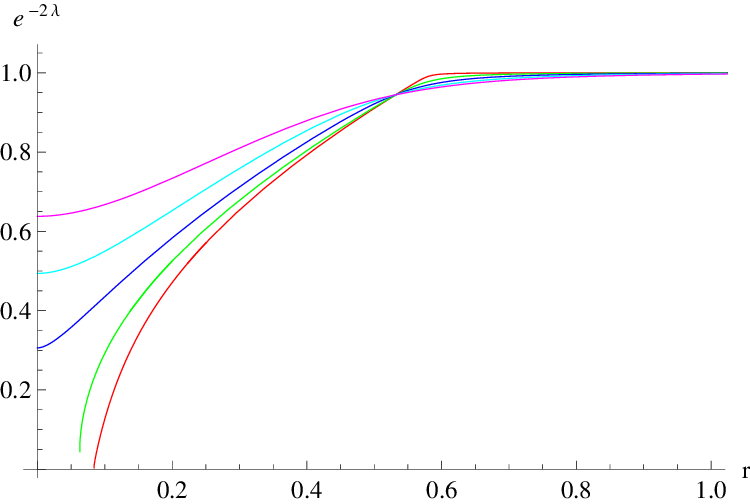}}
\medskip
\centerline{\epsfysize=4.5truecm\epsfbox{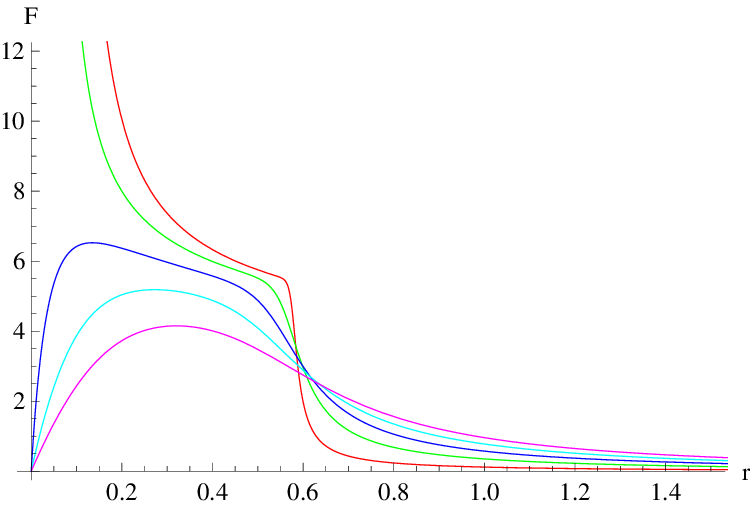}}
\medskip
\baselineskip10pt{\noindent{\smalll Fig.\ 2: The metric function $\scriptstyle{e^{2\n}}$ and  $\scriptstyle{e^{-2\la}}$ and the
electric field $\scriptstyle{F}$ for black holes
with mass $\scriptstyle{M\;=\;1}$, $\scriptstyle{P\;=\;0}$, and $\scriptstyle{Q\;=\;0.1}$ (in cyano), $\scriptstyle{Q\;=\;0.3}$ (in blue),
$\scriptstyle{Q\;=\;0.5}$ (in green), $\scriptstyle{Q\;=\;0.7}$ (in red),  $\scriptstyle{Q\;=\;0.9}$ (in magenta). As it is evident from
the graphs, the curves with  $\scriptstyle{Q\;=\;0.1}$, $\scriptstyle{Q\;=\;0.3}$ display a singularity at $\scriptstyle{r_0\;=\;0.08}$,
$\scriptstyle{r_0\;=\;0.06}$  respectively.}
\bigskip
\vfil\eject
\baselineskip12pt
\section{5. Conclusions}
We have studied the solutions of the dimensionally reduced Kaluza-Klein theory with Einstein-Gauss-Bonnet action, including the
interaction terms that were disregarded in ref.~[7].
A surprising consequence of the inclusion of $L_{int}$ in the action is that dyonic solutions no longer display a regular electric field,
like the solutions of sect.~2 [7], but instead such behaviour can occur for pure electric solutions.

An extension  of our research would be the introduction of a scalar field in the ansatz of dimensional reduction [14,15].
In this case the GB term is present also in the reduced four-dimensional theory, coupled to the scalar.
This fact may modify the structure of the solutions, see e.g.~[16].

\beginref
\ref [1] R. Kerner, Ann. Inst. H. Poincar\'e Phys.Theor. 9, 143 (1968).
\ref [2] R. Kerner, C.\ R.\ Acad.\ Sc.\ Paris {\bf 304}, 621 (1987).
\ref [3] H.A. Buchdal, \JoP{A12}, 1037 (1979).
\ref [4] W. Heisenberg and H. Euler, Z. Phys. {\bf 98}, 714 (1936).
\ref [5] J. Pleba\'nski, {\it Lectures on non-linear electrodynamics}, NORDITA, Copenhagen, 1968.
\ref [6] R. Kerner, in {\it Proceedings of Varna Summer School, "Infinite Dimensional Lie Algebras and Quantum Field Theory"},
World Scientific 1987, \arx{2303.10603}.
\ref [7] S. Mignemi,  \IJMP{A37}, 2250065 (2022).
\ref [8] H.H. Soleng and \O. Gr\o n, \AoP{240}, 432 (1995).
\ref [9] D. Lovelock, \JMP{12}, 498 (1971).
\ref [10] B. Zumino, Phys. Rep. {\bf 137}, 109 (1986).
\ref [11] T. Kaluza, Sitz. Preuss. Akad. Wiss., Math. Phys. {\bf 1}, 966 (1921).
\ref [12] O. Klein, Z. Phys.\ {\bf 37}, 895 (1926).
\ref [13] F. M\"uller-Hoissen, \PL{B201}, 325 (1998).
\ref [14] P. Jordan, Naturwissenschaften {\bf 11}, 250 (1946).
\ref [15] M.Y. Thiry, Compt. Rend. Acad. Sci. Paris {\bf 226}, 216 (1948).
\ref [16] S. Mignemi and N.R. Stewart, \PR{D47}, 5259 (1993).
\end

\ref [2] M. Born and L. lnfeld, \PRS{A144}, 435 (1934).
\ref [4] G.W. Horndeski, \JMP{17}, 1980 (1976).

\ref [12] S.I. Kruglov, \PR{D75}, 117301 (2007).
\ref [13] R. Pellicer and R. J. Torrence, \JMP{19}, 1718 (1969).
\ref [14] H.P. de Oliveira, \CQG{11}, 1469 (1994).
\ref [15] A. Garcia, E. Hackmann, C. L\"ammerzahl and A. Mac\'{\i}as, \PR{D86}, 024037 (2012).
\ref [16] W. Israel, \CMP{8}, 245 (1968).
\ref [17] M. Demia\'nski, Found.\ Phys.\ {\bf 16}, 187 (1986).
\ref [18] G. Gibbons and K. Maeda, \NP{B298}, 741 (1988).
\ref [19] S. Mignemi and N.R. Stewart, \PR{D47}, 5259 (1993).
\ref [20] M.S. Volkov and D.V. Gal'tsov,  JETP Lett.\ {\bf 50} 346 (1989).
\ref [21] J. Schwinger, Science {\bf 165}, 757 (1969).

\ref [23] S.I. Kruglov, Grav.\ and Cosm.\ {\bf 27}, 78 (2021).
\ref [24] P.C.W. Davies, \PRS{A353}, 499 (1977).

\end